\documentclass[12pt]{iopart}

\usepackage{iopams}
\usepackage{pdfsync}
\usepackage{times,dsfont,amssymb,amsthm,amsfonts,amsbsy,amsmath,mathrsfs,bm,bbm,graphicx,graphics,epsfig,multirow,mathtools,color,subfigure,relsize,bbold,mathtools}

\newcommand{\abs}[1]{\lvert#1\rvert}
\newcommand{\bdif}[1]{d^{\,4}\bm{#1}\,}

\begin{document}

\title{Operational Discord Measure for Gaussian States with Gaussian Measurements}

\author{Saleh Rahimi-Keshari$^{1,2}$, Timothy C.~Ralph$^1$, Carlton M.~Caves$^{2,3}$}

\address{$^1$Centre for Quantum Computation and Communication Technology,
School of Mathematics and Physics, University of Queensland, St Lucia, Queensland 4072, Australia}

\address{$^2$Center for Quantum Information and Control, University of New Mexico,
MSC07-4220, Albuquerque, New Mexico 87131-0001, USA}
\address{$^{3}$Centre for Engineered Quantum Systems, School of Mathematics and Physics,
University of Queensland, St Lucia, Queensland 4072, Australia}

\begin{abstract}
We introduce an operational discord-type measure for quantifying nonclassical correlations in bipartite Gaussian states based on using Gaussian measurements. We refer to this measure as operational Gaussian discord (OGD). It is defined as the difference between the entropies of two conditional probability distributions associated to one subsystem, which are obtained by performing optimal local and joint Gaussian measurements. We demonstrate the operational significance of this measure in terms of a Gaussian quantum protocol for extracting maximal information about an encoded classical signal. As examples, we calculate OGD for several Gaussian states in the standard form.
\end{abstract}


\maketitle

\section{Introduction}
\label{sec:intro}

Characterization and quantification of correlations in quantum systems are central in implementing quantum information processing tasks that cannot be done classically. Quantum discord was proposed as a measure of nonclassical correlations, which can capture correlations beyond quantum entanglement~\cite{Zurek01,Zurek00,Vedral01}. This measure of correlation was shown to be useful to characterize resources in a quantum computational model (DQC1)~\cite{DSCaves08}, quantum state merging~\cite{madhok11,cavalcanti}, remote state preparation~\cite{Dakic12}, encoding information onto a quantum state~\cite{Gu12}, quantum phase estimation~\cite{Girolami14}, and quantum key distribution~\cite{Pirandola13}. It was also shown that quantum discord is linked to entanglement generated by the activation protocol~\cite{Piani11} or by a measurement~\cite{Bruss11}.

In general, measures of quantum correlations can be defined as the difference between a quantum entropic measure and a classical entropic measure that is obtained from outcome probabilities of local measurements~\cite{MLang11}. For a bipartite system in quantum state $\rho_{AB}$, quantum discord from subsystem $B$ to $A$ is defined as
\begin{equation}
D(B\rightarrow A)=H^{(\text{min})}_L(\tilde A|\tilde B)-S(A|B),
\end{equation}
where $H^{(\text{min})}_L(\tilde A|\tilde B)$ is the minimized classical conditional entropy, and the quantum conditional entropy is $S(A|B)=S(A,B)-S(B)$, with $S(A,B)$ and $S(B)$ being the von Neumann entropies of the joint state and the marginal state $\rho_B=\text{Tr}_A[\rho_{AB}]$. The classical conditional entropy $H^{(\text{min})}_L(\tilde A|\tilde B)$ is obtained from outcome probability distributions of measurements on $A$ in the eigenbasis of the conditional states $\rho_{A|b}=\text{Tr}_B[\rho_{AB}\Pi_b]/p_b$ obtained after performing a measurement on $B$ described by POVM elements $\{\Pi_b\}$, where $p_b=\text{Tr}_{AB}[\rho_{AB}\Pi_b]$ is the probability for outcome $b$~\cite{MLang11}. Hence, we have
\begin{equation}
\label{con-cla-ent}
H^{(\text{min})}_L(\tilde A|\tilde B)=\min_{\{\Pi_b\}}\sum_{b}p_b S(A|b),
\end{equation}
where $S(A|b)$ is the von Neumann entropy of $\rho_{A|b}$ and the remaining minimization is over the local measurements on $B$. In general, it is not clear how to perform this minimization for an arbitrary quantum state; however, it can be done for certain cases, including a large class of two-qubit states~\cite{Luo}.

Quantum discord was generalized to quantify nonclassical correlations in continuous-variable systems, particularly Gaussian states. Interestingly, all Gaussian states except product states have nonzero quantum discord~\cite{VerDis13,ExpVerDis14}. Gaussian quantum discord (GQD) was introduced as a measure of nonclassical correlations for Gaussian states in which the minimization in the classical conditional entropy~(\ref{con-cla-ent}) is restricted to Gaussian measurements on $B$~\cite{Datta10,Paris10}.  To get to the form~(\ref{con-cla-ent}), however, already requires nonGaussian measurements on the conditional states of $A$; these are generally measurements in a displaced squeezed number basis.  Thus, the GQD cannot really be used as a figure of merit for Gaussian quantum protocols that only involve Gaussian states, Gaussian operations, and Gaussian measurements, such as a Gaussian version of the protocol in~\cite{Gu12}.

By using the optimality of input Gaussian states for Gaussian channels~\cite{Cerf-Holevo,Holevo}, it was recently shown~\cite{Pirandola} that for a
large class of Gaussian states, no nonGaussian measurements on $B$ can further minimize the value of quantum discord, implying that Gaussian quantum discord is equal to quantum discord.  It seems to be an open question whether this is true for all Gaussian states.

In this paper, we introduce a new measure for quantifying nonclassical correlations in bipartite Gaussian states, based solely on Gaussian measurements, which has qualitatively different behavior from GQD.  This measure is defined as the difference between the Gaussian version of the classical conditional entropy $H^{(\text{min})}_{GL}(\tilde A|\tilde B)$, which is given by minimizing over local Gaussian measurements on both subsystems, and the minimum conditional entropy that can be measured by a joint Gaussian measurement, $H^{(\text{min})}_{GJ}(\tilde A|\tilde B)$. We refer to this measure as Operational Gaussian Discord (OGD) because, firstly, it only depends on quantities that can be measured via Gaussian operations and, secondly, it has an operational significance in terms of a quantum protocol that is a Gaussian version of the protocol in~\cite{Gu12}. In this protocol, a classical signal with a Gaussian probability distribution is encoded on one subsystem of a bipartite Gaussian state; using a local Gaussian or a joint Gaussian measurement, one tries to retrieve the signal from the noise associated with the joint state and the measurement.  The optimal measurement is the one that maximizes the classical mutual information between the measurement outcome and the input signal.  We show that in the limit of large variances for the signal probability distribution, the difference between the maximal classical mutual informations obtained by optimal joint and local Gaussian measurements is equal to the OGD of the bipartite Gaussian state.

This paper is structured as follows. In the following section, we review Gaussian quantum discord. We introduce OGD in Sec.~\ref{sec:OGD} and calculate it for several Gaussian states in the standard form in Sec.~\ref{sec:examples}. We demonstrate the operational significance of our measure in Sec.~\ref{sec:Ope-Sig}. We conclude the paper in the last section and pose an open problem. This paper is supplemented with one appendix.

\section{Gaussian quantum discord}
\label{sec:GQD}

Quantum states that have Gaussian Wigner functions are known as Gaussian states.  We gather the phase-space quadratures for a two-mode system into a vector $\hat{\bm{X}}=(\hat{x}_A,\hat{p}_A,\hat{x}_B,\hat{p}_B)$.  A Gaussian state $\rho_{AB}$ can be fully characterized in terms of the mean quadratures $\langle\hat{\bm{X}}\rangle=\bm X$ and the covariance matrix, which has elements $[\sigma_{AB}]_{ij}=\frac12\langle \hat{\bm{X}}_i \hat{\bm{X}}_j+\hat{\bm{X}}_j \hat{\bm{X}}_i\rangle-\langle\hat{\bm{X}}_i\rangle\langle\hat{\bm{X}}_j\rangle$.
As correlations in Gaussian states are independent of local displacement operations, we can, without loss of generality, assume that $\bm{X}=0$. Also, by applying local phase shifts and squeezing operations, any covariance matrix,
\begin{equation}
\bm{\sigma}_{AB}=
\begin{pmatrix}
\mathbf{A} & \mathbf{C} \\
\mathbf{C}^{T} & \mathbf{B}
\end{pmatrix},
\label{Cov-Mat}
\end{equation}
can be brought to the standard form in which $\mathbf{A}=\text{diag}(a,a)$, $\mathbf{B}=\text{diag}(b,b)$, and $\mathbf{C}=\text{diag}(c,d)$~\cite{Adesso-Illuminati,RSimon00}, i.e.,
\begin{equation}\label{standardform}
\bm{\sigma}_{AB}=
\begin{pmatrix}
a&0&c&0\\
0&a&0&d\\
c&0&b&0\\
0&d&0&b
\end{pmatrix}.
\end{equation}
Matrices $\mathbf{A}$ and $\mathbf{B}$ are the covariance matrices of the marginal states $\rho_A=\Tr_B[\rho_{AB}]$ and $\rho_B=\Tr_A[\rho_{AB}]$, and the matrix $\mathbf{C}$ contains the quadrature correlations between the modes. A Gaussian state has zero discord if and only if $\mathbf{C}=0$~\cite{VerDis13}.

A closed-form expression for calculating the classical conditional entropy $H^{(\text{min})}_L(\tilde A|\tilde B)$ with restriction to Gaussian measurements on $B$ was given in Ref.~\cite{Datta10}. A local Gaussian measurement is described by POVM elements that are proportional to pure, single-mode Gaussian states (i.e., rank-one Gaussian operators) with covariance matrix
\begin{equation}
\label{Loc-Cov}
\bm\mu_B=
\begin{pmatrix}
\cos \theta_B & -\sin \theta_B \\
\sin \theta_B & \cos \theta_B
\end{pmatrix}
\begin{pmatrix}
L_B & 0 \\
0 & 1/L_B
\end{pmatrix}
\begin{pmatrix}
\cos \theta_B & \sin \theta_B \\
-\sin \theta_B & \cos \theta_B
\end{pmatrix},
\end{equation}
which is, in fact, the covariance matrix of a squeezed-vacuum state.  The various outcomes of the measurement correspond to the points $\{b\}=\{x_B,p_B\}$ in the phase plane; the corresponding POVM elements are obtained by displacing the squeezed-vacuum state to all points in the phase plane.  (More generally, a single-mode Gaussian measurement can have POVM elements that are Gaussian convex combinations of the rank-one POVM elements, i.e., that are proportional to mixed, single-mode Gaussian states, but such measurements are noisier versions of the rank-one Gaussian measurements and thus are never optimal for our considerations.)   Homodyne measurement has $L_B=0$; heterodyne measurement has $L_B=1$; and for measurements in between, $0<L_B<1$.

After performing such a measurement on $B$ with outcomes $b$, the conditional state $\rho_{A|b}$ has mean quadratures that depend on $b$, but its covariance matrix, given by
\begin{equation}
\label{Con-CM}
\bm\sigma_{A}=\mathbf{A}-\mathbf{C}(\mathbf{B}+\bm\mu_B)^{-1}\mathbf{C}^{T},
\end{equation}
is independent of $b$~\cite{Cirac02}.  Thus, the eigenstates of $\rho_{A|b}$ are, in general, displaced squeezed number states; measuring in this basis minimizes the the Shannon entropy of the outcome probability distribution, making it equal to the von Neumann entropy $S(A|b)$ of the conditional state $\rho_{A|b}$.  This von Neumann entropy is given by $S(A|b)=F\big(\sqrt{\det \bm\sigma_{A}}\big)$~\cite{Adesso-Illuminati}, with
\begin{equation}
F(x)\equiv\frac{x+1}{2}\ln\frac{x+1}{2}-\frac{x-1}{2}\ln\frac{x-1}{2}.
\end{equation}
The Gaussian quantum discord (GQD) is then given by
\begin{equation}
\label{Gau-dis}
D_{\text{GQD}}(B\rightarrow A)=H^{(\text{min})}_L(\tilde A|\tilde B)-S(A|B),
\end{equation}
where the classical conditional entropy,
\begin{equation}\label{HpartG}
H^{(\text{min})}_L(\tilde A|\tilde B)
=\min_{\bm\mu_B}S(A|b)=\min_{\bm\mu_B}F\Big(\sqrt{\det\bm\sigma_{A}}\,\Big),
\end{equation}
is now obtained by minimizing $S(A|b)$ only over Gaussian measurements on~$B$.

For Gaussian states in the standard form~(\ref{standardform}) and having $a=b=c+1$, it is interesting to note that the quantum conditional entropy of the state with $d=c$, referred to as the correlated-correlated (CC) state, is smaller than the quantum conditional entropy of the state with $d=-c$, referred to as the correlated-anticorrelated (CA) state. On the other hand, the classical conditional entropy~(\ref{HpartG}) is the same for these two separable states~\cite{Datta10}. This implies that the GQD of the CC state is larger than that of the CA state, although the marginal states of these two separable states are the same.  Note that given $c$, the CC and CA states are the nonentangled states that have maximal correlations in their quadratures.

Recently, using a connection between the continuous (differential) Shannon entropy of the Wigner function and the R\'enyi-2 entropy, Gaussian R\'enyi-2 discord was defined as a measure of nonclassical correlations in Gaussian states~\cite{Adesso12}. In this measure, the von Neumann entropies in Eq.~(\ref{Gau-dis}) are replaced by R\'enyi-2 entropies
\begin{equation}
\label{Gau-Ren-dis}
D_2(B\rightarrow A)=\min_{\bm\mu_B}S_2(A|b)-S_2(A|B),
\end{equation}
where $S_2(A|b)=-\ln(\Tr[\rho_{A|b}^2])=\frac12\ln(\det\bm\sigma_{A})$ and $S_2(A|B)=\frac12\ln(\det \bm{\sigma}_{AB}/\det \mathbf{B})$. Notice that, the conditional entropy $S_2(A|b)$ corresponds to the continuous Shannon entropy of the Wigner function of $\rho_{A|b}$ up to a constant~\cite{Adesso12}.  There is, however, no Gaussian measurement whose outcome probability distribution is equal to the Wigner function, as the noncommuting observables $\hat{x}_A$ and $\hat{p}_A$ cannot be measured simultaneously without some noise penalty.  The CC and CA states have the same Gaussian R\'enyi-2 discord, and Gaussian states with no correlations in one of the quadratures ($d=0$) have zero Gaussian R\'enyi-2 discord.

Neither the GQD nor the Gaussian R\'enyi-2 discord satisfy the condition of nonclassical correlations~\cite{MLang11} for Gaussian protocols, because they use a nonGaussian measurement on~$A$.  We turn now to formulating an operational, discord-type measure of nonclassical correlations for Gaussian states that is based purely on Gaussian measurements.

\section{Operational Gaussian discord}
\label{sec:OGD}

We refer to our new measure as operational Gaussian discord (OGD) and  define it as
\begin{equation}
\label{Gau-Dis-Typ}
D_{\text{OGD}}(B\rightarrow A)=H^{(\text{min})}_{GL}(\tilde A|\tilde B)-H^{(\text{min})}_{GJ}(\tilde A|\tilde B),
\end{equation}
where $H^{(\text{min})}_{GL}(\tilde A|\tilde B)$ is the minimum conditional entropy of $A$ after performing \emph{local\/} Gaussian measurements on $A$ and $B$, and $H^{(\text{min})}_{GJ}(\tilde A|\tilde B)$ is the minimum conditional entropy of the same subsystem after performing a \emph{joint\/} Gaussian measurement on $A$ and $B$. The entropies are the continuous (differential) Shannon entropy of Gaussian probability distributions, which for a single mode are given by $\frac12\ln(\det \tilde{\bm{\sigma}})+\ln(2\pi e)$, with $\tilde{\bm{\sigma}}$ being the covariance matrix of the probability distribution (see Appendix~A). In our notation, $\tilde A$ and $\tilde B$ denote that the entropies are calculated using outcome probability distributions of the measurements. As all the probability distributions are Gaussian, in order to calculate the OGD~(\ref{Gau-Dis-Typ}), one just needs to minimize the determinants of the covariance matrices of the conditional Gaussian probability distributions for the outcomes of local and joint Gaussian measurements. For a discussion of Gaussian measurements, conditional probability distributions, and the corresponding entropies, see Appendix~A.

In general, the POVM elements of a two-mode Gaussian measurement are proportional to two-mode Gaussian states whose covariance matrix, according to the Williamson theorem~\cite{Williamson}, can be written as
\begin{equation}
\label{Gen-Cov}
\bm\mu_J= \mathbf{S}^T( \nu_1 \mathbb{1} \oplus \nu_2 \mathbb{1}) \mathbf{S}.
\end{equation}
Here the $2\times2$ identity matrix $\mathbb{1}$ represents the single-mode vacuum state (the choice of units can be thought of as setting $\hbar=2$), $\nu_1\mathbb{1} \oplus \nu_2 \mathbb{1}$ corresponds to product thermal states with variances $\nu_1$ and $\nu_2$ in the two modes, and $\mathbf{S}$ represents a symplectic transformation~\cite{Weedbrook}.  For our purpose, that is to minimize the entropies of the outcome probability distributions, we consider Gaussian measurements that have rank-one POVM elements, i.e., $\nu_1 =\nu_2=1$; thus, the POVM elements are proportional to pure Gaussian states.  Measurements with mixed POVMs will add more noise and increase the entropy. Any symplectic matrix can be expressed as $\mathbf{S}= \mathbf{K} [\mathbf{s}(r_1)\oplus \mathbf{s}(r_2)] \mathbf{L}$, where $\mathbf{K}$ and $\mathbf{L}$ represent beamsplitter transformations and $\mathbf{s}(r)$ represents a single-mode squeezing operation~\cite{Euler-decom,Braunstein}.
Using this expression for $\mathbf{S}$ in Eq.~(\ref{Gen-Cov}) and knowing that the action of a beamsplitter on vacuum states results in vacuum states, we can write the covariance matrix of the POVM elements of a two-mode, joint Gaussian measurement in the form
\begin{align}
\label{Joi-Cov}
\bm\mu_J=\mathbf{R}^T(\phi_A,\phi_B)\mathbf{B}^T(\eta)(\bm\mu_A \oplus \bm\mu_B)\mathbf{B}(\eta)\mathbf{R}(\phi_A,\phi_B)
=\begin{pmatrix}
\bm\mu_{A,J} & \bm\mu_{C,J}  \\
\bm\mu_{C,J}^T  & \bm\mu_{B,J}
\end{pmatrix},
\end{align}
where
\begin{equation}
B(\eta)=
\begin{pmatrix}
\sqrt{\eta} & 0 & -\sqrt{1-\eta} & 0 \\
0 & \sqrt{\eta} & 0 & -\sqrt{1-\eta} \\
\sqrt{1-\eta} & 0 & \sqrt{\eta} & 0 \\
0 & \sqrt{1-\eta} & 0 & \sqrt{\eta}
\end{pmatrix}
\end{equation}
describes a beamsplitter transformation,
\begin{equation}
R(\phi_A,\phi_B)=
\begin{pmatrix}
\cos \phi_A & -\sin \phi_A & 0 & 0 \\
\sin \phi_A & \cos \phi_A & 0 & 0 \\
0 & 0 & \cos \phi_B & -\sin \phi_B \\
0 & 0 & \sin \phi_B & \cos \phi_B
\end{pmatrix}
\end{equation}
describes pre-beamsplitter single-mode phase shifts, and $\bm\mu_B$ is defined as in Eq.~(\ref{Loc-Cov}), with
$\bm\mu_A$ defined analogously.  As we see from the above expression, the joint Gaussian measurement corresponding to this covariance matrix can be realized by two phase shifters and a beamsplitter followed by two local Gaussian measurements.  Obviously, for $\phi_A=\phi_B=0$ and $\eta=1$, we obtain the covariance matrix of a local Gaussian measurement,
\begin{equation}
\label{Loc-Cov-2}
\bm\mu_L=\bm\mu_A \oplus \bm\mu_B.
\end{equation}

As shown in the appendix, after performing a joint Gaussian measurement, the covariance matrix of the conditional probability distribution for $A$ is given by
\begin{equation}
\label{Con-CM-JGM}
\tilde{\bm\sigma}_{A,J}=\tilde{\mathbf{A}}-\tilde{\mathbf{C}} \tilde{\mathbf{B}}^{-1}\tilde{\mathbf{C}}^{T},
\end{equation}
which is obtained from a joint Gaussian probability distribution with the covariance matrix
\begin{equation}
\label{CM-JM}
\tilde{\bm\sigma}_{AB,J}=\bm\sigma_{AB}+\bm\mu_J=
\begin{pmatrix}
\mathbf{A}+\bm\mu_{A,J} & \mathbf{C}+\bm\mu_{C,J} \\
\mathbf{C}^{T}+\bm\mu_{C,J}^T & \mathbf{B}+\bm\mu_{B,J}
\end{pmatrix}
=\begin{pmatrix}
\tilde{\mathbf{A}}&\tilde{\mathbf{C}} \\
\tilde{\mathbf{C}}^{T}&\tilde{\mathbf{B}}
\end{pmatrix}.
\end{equation}
After local Gaussian measurements on $A$ and $B$, the covariance matrix of the conditional probability distribution for $A$ is
\begin{equation}
\label{Con-CM-LGM}
\tilde{\bm\sigma}_{A,L}=\mathbf{A}+ \bm\mu_A-\mathbf{C}(\mathbf{B}+\bm\mu_B)^{-1}\mathbf{C}^{T}.
\end{equation}
Thus, by using Eqs.~(\ref{Con-CM-LGM}) and~(\ref{Con-CM-JGM}), the OGD measure becomes
\begin{align}
\label{Gau-Dis-Typ-2}
D_{\text{OGD}}(B\rightarrow A)= \min_{\bm\mu_A,\bm\mu_B}
\frac{1}{2}
\ln\left(\det\tilde{\bm\sigma}_{A,L}\right)-
\min_{\bm\mu_J}\frac{1}{2}\ln\left(\det\tilde{\bm\sigma}_{A,J}\right).
\end{align}

Operational Gaussian discord is always nonnegative, because the set of all joint measurements includes all local measurements; hence, the conditional entropy minimized over all possible joint measurements, $H_{GJ}^{\rm min}(\tilde A|\tilde B)$, can never be larger than the conditional entropy miminized over all local measurements, $H_{GL}^{\rm min}(\tilde A|\tilde B)$. In addition, the OGD of product states is zero. In this case, for a joint measurement we have $\tilde{\bm\sigma}_{A,J}=\mathbf{A}+\bm\mu_{A}^{\prime}$ with $\bm\mu_{A}^{\prime}=\bm\mu_{A,J}-\bm\mu_{C,J}(\mathbf{B}+\bm\mu_{B,J})^{-1}\bm\mu_{C,J}^{T}$, which is equivalent to a local measurement on $A$ with covariance matrix $\bm\mu'_{A}$; hence, $\det \tilde{\bm\sigma}_{A,J}$ cannot be smaller than $\det \tilde{\bm\sigma}_{A,L}$.

\section{Examples: operational Gaussian discord for some Gaussian states}
\label{sec:examples}

In general, it is not clear how to obtain a closed-form expression for OGD for an arbitrary Gaussian quantum state. The conditional entropy with local Gaussian measurements must be minimized over four parameters, $\{\theta_A,\theta_B,L_A,L_B\}$, and the conditional entropy with joint Gaussian measurements must be minimized over seven parameters, $\{\phi_A,\phi_B,\eta,\theta_A,\theta_B,L_A,L_B\}$. In the following, by using analytical and numerical methods, we calculate OGD for some Gaussian states in the standard form~(\ref{standardform}).

\subsection{Entangled and separable Gaussian states}

Let us consider a class of Gaussian states whose covariance matrices~(\ref{standardform}) are parameterized by $a$ and $t$ such that $a=b\geq1$ and $c=-d=t\sqrt{a^2-1}\geq 0$, where $0\leq t\leq1$.  When $t=1$, this is a pure two-mode squeezed-vacuum state, with $a=b=\cosh 2r$ and $c=-d=\sinh 2r$, $r$ being the squeezing parameter.  For $t>\sqrt{(a-1)/(a+1)}$, the state is entangled~\cite{RSimon00}; for the other values of $t$, the state is separable.  The boundary between separability and entanglement, i.e., $t=\sqrt{(a-1)/(a+1)}$, is occupied by the CA state.

By minimizing $\det \tilde{\bm\sigma}_{A,L}$ over all local Gaussian measurements we find that the optimal local measurements for all values of $a$ and $t$ are two heterodyne measurements, i.e., $L_A=L_B=1$. This gives a symmetric covariance matrix for the conditional probability distribution:
\begin{equation}
\label{Con-CM-LGM-t-a}
\tilde{\bm\sigma}_{A,L}=
\begin{pmatrix}
1+a+(1-a)t^2 & 0 \\
0 & 1+a+(1-a)t^2
\end{pmatrix}.
\end{equation}

In order to minimize $\det \tilde{\bm\sigma}_{A,J}$ one can guess that, as the quadratures are equally correlated but with a different sign, the covariance matrix of the POVM elements of the joint Gaussian measurement must be in the same form, with the sum $\bm\sigma_{AB}+\bm\mu_J=\tilde{\bm\sigma}_{AB,J}$ enhancing the correlations and minimizing the determinant of the conditional covariance matrix $\tilde{\bm\sigma}_{A,J}$. This means that $\bm\mu_J$ must be the covariance matrix of a two-mode squeezed state, as it is the only pure Gaussian state in that form,
\begin{align}
\label{2modsq-CM1}
\bm\mu_{A,J}&=\bm\mu_{B,J}=
\begin{pmatrix}
\frac12(1/L+L) & 0  \\
0  & \frac12(1/L+L)
\end{pmatrix},\\[4pt]
\label{2modsq-CM2}
\bm\mu_{C,J}&=
\begin{pmatrix}
\frac12(1/L-L) & 0  \\
0  & -\frac12(1/L-L)
\end{pmatrix},
\end{align}
i.e., $\phi_A=\phi_B=\theta_A=0$, $\theta_B=\pi/2$, $\eta=0.5$, and $L_A=L_B=L$. Numerical calculations confirm that this is the optimal choice for the joint Gaussian measurement. Minimizing $\det\tilde{\bm{\sigma}}_{A,J}$ over the parameter $L$ gives
\begin{equation}
L=
\begin{cases}
    \displaystyle{\frac{1-a t^2-a-t^2+2t \sqrt{a^2-1}}{1+t^2-a \left(1-t^2\right)}},
    &0\leq t<\sqrt{(a-1)/(a+1)}, \\[3pt]
    0,
    & \sqrt{(a-1)/(a+1)}\leq t\leq1.
\end{cases}
\end{equation}
The corresponding covariance matrix of the conditional probability distribution for $0\leq t<\sqrt{(a-1)/(a+1)}$ is
\begin{equation}
\tilde{\bm{\sigma}}_{A,J}=
\left(
\begin{array}{cc}
 (1+a) \left(1-t^2\right) & 0 \\
 0 & (1+a) \left(1-t^2\right)
\end{array}
\right),
\end{equation}
and for other values of $t$ is
\begin{equation}
\tilde{\bm{\sigma}}_{A,J}=
2\left(
\begin{array}{cc}
  a-t\sqrt{a^2-1}  & 0 \\
 0 &  a-t\sqrt{a^2-1}
\end{array}
\right).
\end{equation}
For the CA state and entangled states, an optimal measurement is a beamsplitter followed by two homodyne measurements (POVM elements are two-mode infinitely squeezed states).  For the separable states, the local measurements after the beamsplitter are measurements in a displaced squeezed-vacuum basis, varying between heterodyne for $t=0$ and homodyne for $t=\sqrt{(a-1)/(a+1)}$.

The OGD for these states is given by
\begin{align}
\begin{split}
D_{\text{OGD}}(B\rightarrow A)&=\ln(1+a+t^2-at^2)\\
&\qquad-
\begin{cases}
    \ln\big((1+a)(1-t^2)\big),&0\leq t<\sqrt{(a-1)/(a+1)}, \\
    \ln(2a-2t\sqrt{a^2-1}),   &\sqrt{(a-1)/(a+1)}\leq t\leq1.
\end{cases}
\end{split}
\end{align}

\begin{figure}[t]
\centering
	\includegraphics[width=0.6\columnwidth]{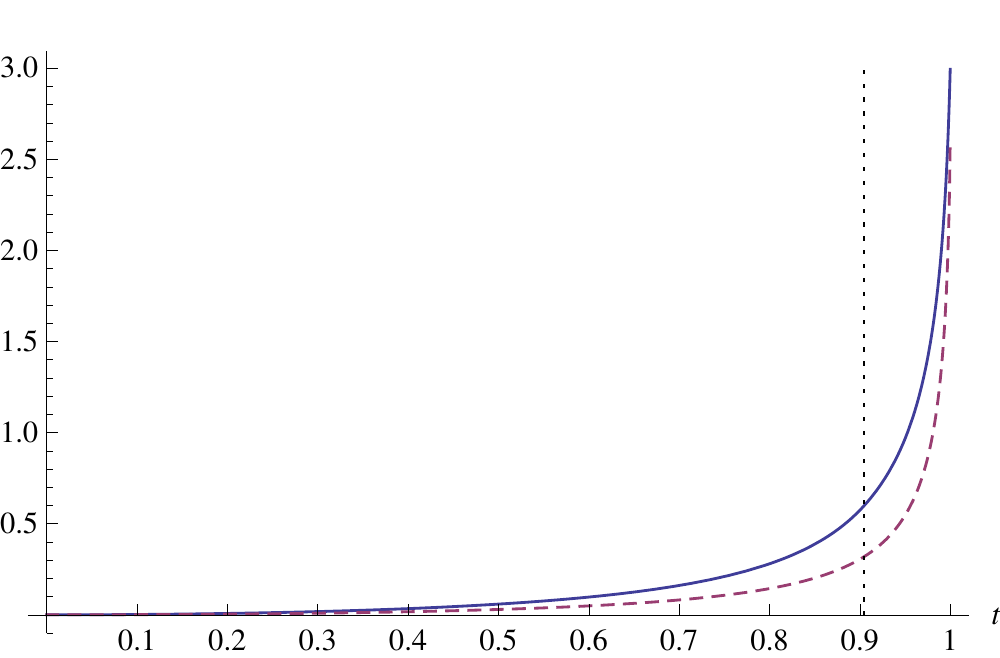}
	\caption{ The solid blue line shows the operational Gaussian discord (OGD), and the dashed red line shows Gaussian quantum discord (GQD) for a Gaussian state in the standard form~(\ref{standardform}), with $a=b=10$ and $c=-d=t\sqrt{99}$.  The optimal local Gaussian measurements are two heterodyne measurements for all values of $t$. For $t\geq\sqrt{9/11}$, the optimal joint Gaussian measurement is a 50:50 beamsplitter followed by two homodyne measurements.  Notice that for $t >\sqrt{9/11}$ (vertical line), the state is entangled.  For other values of $t$, the optimal joint Gaussian measurement is local measurements in a displaced squeezed-vacuum basis after the beamsplitter, varying between no squeezing (heterodyne) for $t=0$ to infinite squeezing (homodyne) for $t=\sqrt{9/11}$.}
	\label{Sep-Ent}
\end{figure}

For these states $D_{\text{GQD}}(B\rightarrow A)\leq D_{\text{OGD}}(B\rightarrow A)$, which we illustrate in a particular case in Fig.~\ref{Sep-Ent}.  This implies that the difference between the classical and quantum conditional entropies is less than or equal to the difference between the conditional entropies obtained by local and joint Gaussian measurements. Also, for the two-mode squeezed-vacuum state, we have $D_{\text{OGD}}(B\rightarrow A)=2r$; the quantum discord of this state is equal to the von Neumann entropy of the marginal state, $S(B)$, which for large values of $r$ scales as $2r+1-2\ln2$.

\subsection{Correlated-correlated and correlated-anticorrelated states}

Here we consider separable Gaussian states parameterized by $c$ and $q$ such that $a=b=c+1$ and $d=q c$, with $c>0$ and $-1\leq q\leq 1$.  The parameter $q$ controls the correlation in the $p$-quadratures; by changing $q$ from $-1$ to $1$, the state changes from the CA state to the CC state.

We first minimize $\det\tilde{\bm{\sigma}}_{A,L}$ for the minimum conditional entropy with a local Gaussian measurement. Numerical calculations show that $\theta_A=\theta_B=0$, as expected because the state is in standard form.  The minimizing values of $L_A$ and $L_B$ are
\begin{equation}
L_A=\sqrt{\frac{2}{2+c-cq^2}}
\end{equation}
and
\begin{equation}
L_B=
\begin{cases}
    \displaystyle{\frac{(1+c)(q^2-1)+c|q|\sqrt{4+2c-2cq^2}}{(1+c)^2-(1+c^2)q^2}},
    &(1+2c)^{-1/2}<\abs{q}\leq1,\\[3pt]
    0,
    & 0\leq \abs{q}\leq(1+2c)^{-1/2}.
\end{cases}
\end{equation}
and these give
\begin{equation}
\det\tilde{\bm\sigma}_{A,L}=\left(1+{L_A}+\frac{c (1+L_B)}{1+c+{L_B}}\right)\left(1+c+\frac{1}{{L_A}}-\frac{c^2 q^2 L_B}{(1+c)L_B +1}\right).
\end{equation}
According to the above expressions, the optimal local measurement on $A$ for all values of $q$ is in a displaced squeezed-vacuum basis, which limits to a heterodyne measurement when $|q|=1$.  The optimal local measurement on $B$ for $(1+2c)^{-1/2}<|q|<1$ is also in a displaced squeezed-vacuum basis, but for the small correlations in the $p$-quadratures, $|q|\leq(1+2c)^{-1/2}$, the optimal local measurement is homodyne.  For $|q|=1$ the local measurements on both $A$ and $B$ are heterodyne measurements; in this case, we have $\sqrt{\det\tilde{\bm\sigma}_{A,L}}=4(1+c)/(2+c)$.

Using numerical calculations, we find that the POVM elements of the optimal joint Gaussian measurement are two-mode squeezed states, with covariance matrix $\bm\mu_J$ given by Eqs.~(\ref{2modsq-CM1}) and (\ref{2modsq-CM2}). We obtain
\begin{equation}
\det \tilde{\bm\sigma}_{A,J}= \frac{4 (1+L) (1+L+2 c L) (1+L+c L-c q L) (1+c+L+c q)}{(1+2 L+2 c L+L^2)^2},
\end{equation}
which is minimized by
\begin{equation}
L=\frac{q-1+\sqrt{4+4c-4c q^2}}{3+2 c (1-q)-q}.
\end{equation}
The expression for $L$ shows that for $q=1$ we have $L=1$; i.e., the optimal joint Gaussian measurement is a 50:50 beamsplitter followed by two heterodyne measurements.  In this case, $\bm\mu_J=\mathbb{1}\oplus\mathbb{1}$, and this measurement is equivalent to two local heterodyne measurements. Moreover, it is easy to see that for $(1+2c)^{-1/2}\leq q\leq1$, the minimum conditional entropies with local and joint Gaussian measurements are the same, $\det\tilde{\bm\sigma}_{A,J}=\det \tilde{\bm\sigma}_{A,L}$, and thus OGD is zero. We also observe that the parameter $L$ decreases as $q$ decreases.  For $q=-1$ we have $L=0$ and $\sqrt{\det \tilde{\bm{\sigma}}_{A,J}}=2$, which corresponds to performing two homodyne measurements, with $\theta_A=0$ and $\theta_B=\pi/2$, after the beamsplitter.

\begin{figure}[t]
\centering
	\includegraphics[width=0.7\columnwidth]{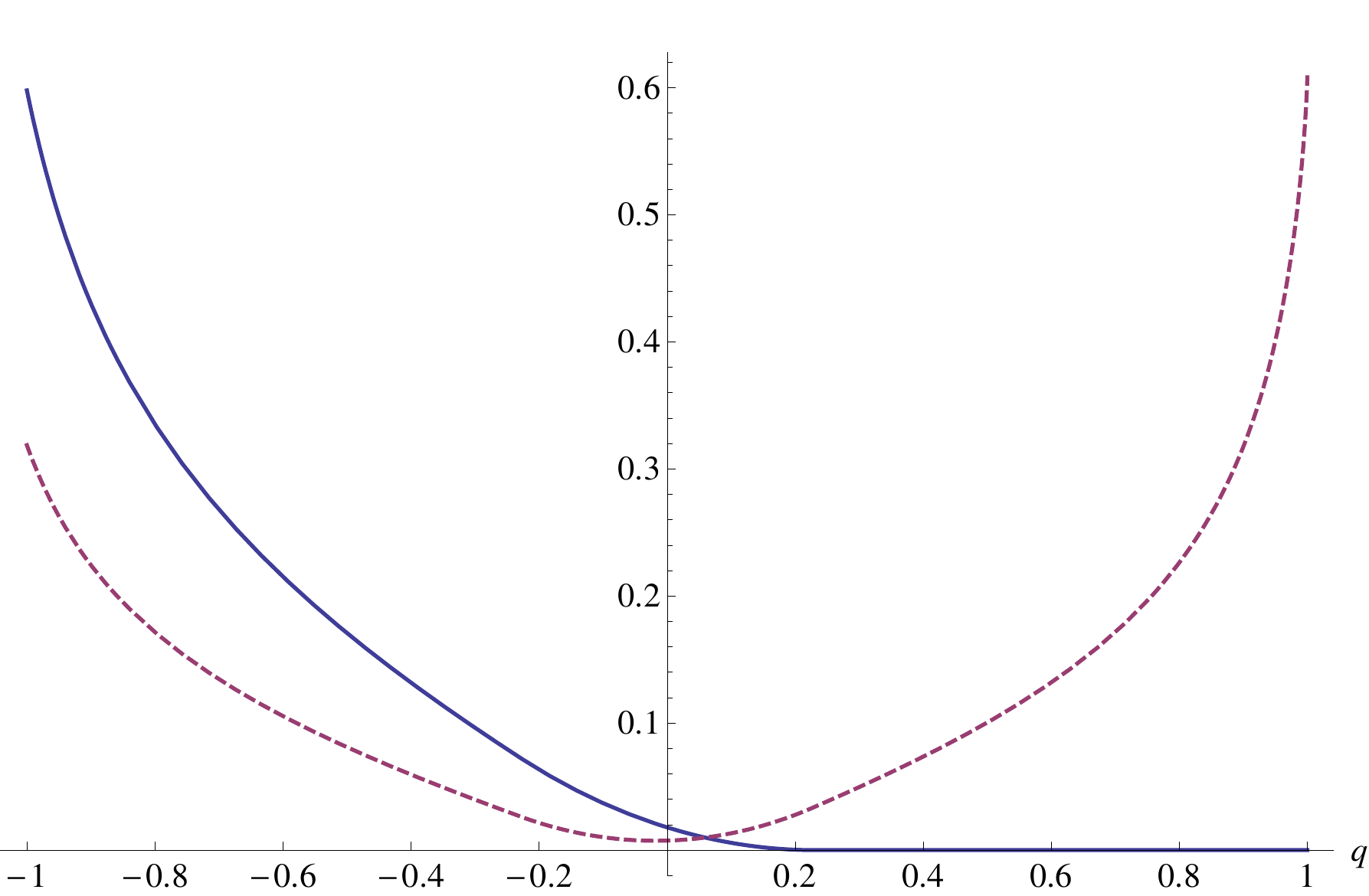}
	\caption{ Operational Gaussian discord (OGD) (solid blue line) and Gaussian quantum discord (GQD) (dashed red line) for Gaussian states with $a=b=c+1=10$ and $d=9q$.  The parameter $q$ controls the correlation between the $p$-quadratures of the joint system.  OGD monotonically decreases in the interval $-1\leq q <1/\sqrt{19}$, and for $1/\sqrt{19}\leq q\leq1$ OGD is zero; note
that $q=1/\sqrt{19}$ is the point at which the optimal local measurement on $B$ changes to homodyne measurement. According to GQD, the CC state, $q=1$, has more nonclassical correlation than the CA state, $q=-1$, but the OGD measure attributes zero nonclassical correlation to the CC state and the maximal nonclassical correlation in this class to the CA state.}
	\label{CC-CA-a10}
\end{figure}

In Fig.~\ref{CC-CA-a10}, we compare GQD and OGD measures for Gaussian states with $c=9$.  As shown, GQD for the state with $q=1$ is larger than for the state with $q=-1$.  In contrast, according to OGD, the state with $q=1$ has zero correlation, and the state with $q=-1$ has the maximum correlation.

\subsection{Asymmetric Gaussian states}
\label{Asy-Sta}

Bipartite Gaussian states whose marginal states are not the same are asymmetric.  To explore properties of such states, we consider separable, asymmetric Gaussian states in the standard form, which are parametrized by $b$, $v$, and $s$ with $a=b+v$, $c=\abs{s}$, and $d=s$, where $b\geq1$, $v\geq0$ and $c=|s|\leq b-1$.

Using numerical and analytical calculations, we find that the optimal local measurements that minimize $\det \tilde{\bm\sigma}_{A,L}$ for all values of $s$ and $v$ are heterodyne measurements, $L_A=L_B=1$, which yields a symmetric covariance matrix for the conditional probability distribution,
\begin{equation}
\label{Asy-Loc-CM}
\tilde{\bm\sigma}_{A,L}=
\begin{pmatrix}
1+b+v-s^2/(1+b) & 0 \\
 0 & 1+b+v-s^2/(1+b)
\end{pmatrix}.
\end{equation}

In order to minimize $\det \tilde{\bm\sigma}_{A,J}$, we consider the cases $s>0$ ($c=d$) and $s<0$ ($c=-d$) separately. Note that for $s=0$ the state is a product state. For $c=d$, as for the CC state in the previous subsection, we find that for all values of $\abs{s}$ and $v$ the optimal joint Gaussian measurement is a 50:50 beamsplitter followed by two heterodyne measurements ($\phi_A=\phi_B=\theta_A=\theta_B=0$, $L_A=L_B=1$, $\eta=1/2$), which implies $\tilde{\bm\sigma}_{A,J}=\tilde{\bm\sigma}_{A,L}$.  For the case $c=-d$, however, we find that the optimal joint Gaussian measurement is described by parameters $\phi_A=\phi_B=\theta_A=0$, $\theta_B=\pi/2$, $\eta=1/2$, and $L_A=L_B=(b-1+s)/(b-1-s)$; i.e., the POVM elements are two-mode squeezed states. In this case, we obtain a symmetric covariance matrix for the conditional probability distribution as
\begin{equation}
\label{Asy-Joi-CM}
\tilde{\bm\sigma}_{A,J}=
\begin{pmatrix}
1+b+v -s^2/(b-1)& 0 \\
 0 & 1+b+v -s^2/(b-1)
\end{pmatrix}.
\end{equation}

As a consequence, OGD for these states is given by
\begin{equation}
D_{\text{OGD}}(B\rightarrow A)=
\begin{cases}
    0,&0\leq s\leq b-1. \\[3pt]
    \displaystyle{\ln\!\left(1 + \frac{2 s^2}{(1 + b)(b^2 - s^2-1-v+bv)}\right)},& 1-b\leq s<0.
\end{cases}
\end{equation}
Notice that the optimal local  and joint Gaussian measurement strategies are independent of the value of $v$, and for $v\rightarrow\infty$, OGD is zero.

\section{Operational significance}
\label{sec:Ope-Sig}

We now present the operational significance of our measure in terms of a Gaussian protocol for encoding information onto Gaussian quantum states (see Fig.~1). In this protocol, two independent classical random variables, $x_s$ and $p_s$, represented by the vector $\bm X_s=(x_s,p_s)$ and described by Gaussian probability distributions with the same variance $V_s$, are encoded on the $x$- and $p$-quadratures of subsystem $A$ of a joint system in the Gaussian state $\rho_{AB}$. The encoding procedure is done by applying the displacement operator $D_{A}(\bm X_s)=\exp[i(p_s\hat{x}_A-x_s\hat{p}_A)/2]$ and averaging over the Gaussian distributions for $x_s$ and $p_s$.  The state after encoding thus becomes
\begin{equation}
\rho'_{AB}=\int d\bm X_s\,\frac{e^{-(x_s^2+p_s^2)/2V_s}}{2\pi V_s} D_{A}(\bm X_s)\rho_{AB} D^{\dagger}_{A}(\bm X_s).
\end{equation}
The state $\rho'_{AB}$ is also Gaussian, with covariance matrix ${\bm\sigma}_{AB}^{\prime}={\bm\sigma}_{AB}+V_s\mathbb{1}\oplus\mathbf{0}$, where $\mathbf{0}$ is the $2\times$2 zero matrix.

The aim is to obtain an estimate of the signals, $\bm Y_e=(x_e,p_e)$, by using some measurement strategy that takes advantage of the correlations between the subsystems in such a way that the classical mutual information $I(\bm X_s,\bm Y_e)$ is maximized.  It was shown, using Holevo's theorem, that for maximal encoding, i.e., $V_s\rightarrow\infty$, the difference between the maximum extractable information with and without restricting to local measurements is equal to quantum discord of the state $\rho_{AB}$~\cite{Gu12}.  In order to saturate the extractable information, however, nonGaussian measurements are required, in the way we described earlier for quantum discord.  Here we consider a Gaussian version of the protocol, in which there are two measurement strategies: local Gaussian measurements and joint Gaussian measurements.

\begin{figure*}
\includegraphics[width=1\textwidth]{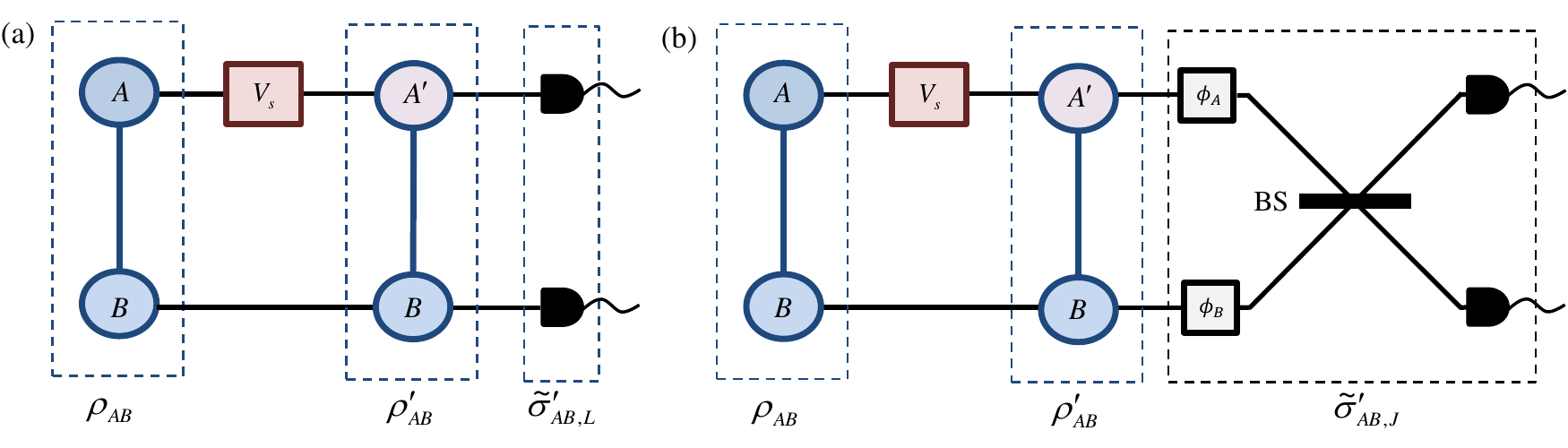}
\caption{By applying a Gaussianly distributed local displacement operator, a pair of classical Gaussian random variables $\bm X_s=(x_s,p_s)$, both having the same variance $V_s$, are encoded on $x$- and $p$-quadratures of subsystem $A$, which is part of a bipartite system in Gaussian state with the covariance matrix $\bm\sigma_{AB}$~\cite{Gu12}.  (a)~In the first strategy, one performs optimal local Gaussian measurements, whose POVM elements are described by the covariance matrix $\bm\mu_A\oplus\bm\mu_B$.   After post-processing the data, one obtains a signal estimate $\bm Y_e=(x_e,p_e)$ such that the mutual information $I_L(\bm X_s,\bm Y_e)$ is maximized.  The covariance matrix of the joint probability distribution is $\tilde{\bm\sigma}_{AB,L}^{\prime}=\bm\sigma_{AB}+\bm\mu_A\oplus\bm\mu_B$.
(b)~In the second strategy, one performs an optimal joint Gaussian measurement such that the mutual information $I_J(\bm X_s,\bm Y_e)$ is maximized.  As shown in the text, the most general form of a joint Gaussian measurement consists of two phase shifters and a beamsplitter (BS) followed by two local Gaussian measurements.  The covariance matrix of the probability distribution for the outcomes of this measurement is $\tilde{\bm\sigma}_{AB,J}^{\prime}=\bm\sigma_{AB}+\bm\mu_J$, where $\bm\mu_J$ is the covariance matrix of the POVM elements.  In the limit of maximal encoding ($V_s\rightarrow\infty$), the difference between $I_J(\bm X_s,\bm Y_e)$ and $I_L(\bm X_s,\bm Y_e)$ is equal to operational Gaussian discord (OGD) of the state $\rho_{AB}$.}\label{local-joint}
\end{figure*}

Consider first the case where subsystem $B$ is not available to us. Assuming the state $\rho_{AB}$ was in the standard form, the marginal state $\rho_A$ is symmetric with variances $a$ in the $x$- and $p$-quadratures. In this case, the covariance matrix of the outcome probability distribution of a Gaussian measurement is given by
\begin{equation}
\begin{pmatrix}
a+L_A+V_s& 0 \\
 0 & a+1/L_A+V_s
\end{pmatrix}.
\end{equation}
By using the expression for the mutual information of two parallel Gaussian channels~\cite{InfThe}, the mutual information between $\bm X_s$ and an estimate of it, $\bm Y_e$, that is obtained after the measurement is given by
\begin{equation}
I(\bm X_s,\bm Y_e)
=\frac{1}{2}\ln\!\left(1+\frac{V_s}{a+L_A}\right)
+\frac{1}{2}\ln\!\left(1+\frac{V_s}{a+1/L_A}\right).
\end{equation}
This quantity is maximized for $L_A=1$, i.e., by performing heterodyne measurement. There are two sources of noise reducing the mutual information: the noise of the quantum state and the noise associated with the measurement.  While the former noise is inevitable due to the uncertainty principle, the latter could be reduced if subsystem $B$ was available to us. In that case, one could take advantage of the correlations by performing some measurement on $B$ and post-processing the outcomes in order to effectively reduce the noise of the state, thus allowing extraction of more information about the signals.

When both subsystems are available, in the first strategy, one performs local Gaussian measurements on subsystems $B$ and $A$. This yields a conditional probability distribution for $A$, with covariance matrix $\tilde{\bm\sigma}_{A,L}^{\prime}=\tilde{\bm\sigma}_{A,L}+V_s\mathbb{1}$,
where $\tilde{\bm\sigma}_{A,L}$ is given by Eq.~(\ref{Con-CM-LGM}). The mutual information, given by
$I(\bm X_s,\bm Y_e)=\frac{1}{2}[\ln(1+V_s/a_{L1})+\ln(1+V_s/a_{L2})]$,
where $a_{L1}$ and $a_{L2}$ are the eigenvalues of $\tilde{\bm\sigma}_{A,L}$, should be maximized over the local measurements, which means to maximize it over the local-measurement covariance matrices $\bm\mu_A$ and $\bm\mu_B$. Thus, the quantity of interest is
\begin{equation}
I_L(\bm X_s,\bm Y_e)
=\max_{\bm\mu_A,\bm\mu_B}
\frac{1}{2}\left[\ln\!\left(1+\frac{V_s}{a_{L1}}\right)
+\ln\!\left(1+\frac{V_s}{a_{L2}}\right)\right].
\end{equation}

In the second strategy, by using a joint Gaussian measurement, one obtains a conditional probability distribution with covariance matrix $\tilde{\bm\sigma}_{A,J}^{\prime}=\tilde{\bm\sigma}_{A,J}+V_s\mathbb{1}$,
where $\tilde{\bm\sigma}_{A,J}$ is given by Eq.~(\ref{Con-CM-JGM}).  The mutual information, $I(\bm X_s,\bm Y_e)=\frac{1}{2}[\ln(1+V_s/a_{J1})+\ln(1+V_s/a_{J2})]$, where $a_{J1}$ and $a_{J2}$ are the eigenvalues of $\tilde{\bm\sigma}_{A,J}$, should be maximized over the covariance matrix $\bm\mu_J$ that describes the joint Gaussian measurement, so the quantity of interest is
\begin{equation}
I_J(\bm X_s,\bm Y_e)
=\max_{\bm\mu_J}
\frac{1}{2}\left[\ln\!\left(1+\frac{V_s}{a_{J1}}\right)
+\ln\!\left(1+\frac{V_s}{a_{J2}}\right)\right].
\end{equation}

In the limit that the classical signals have very large power, $V_s$ is much larger than any of the eigenvalues.  In this situation, we have
\begin{align}
I_L(\bm X_s,\bm Y_e)\simeq
\ln V_s
-\min_{\bm\mu_A,\bm\mu_B}\frac{1}{2}\ln(a_{L1}a_{L2})
=\ln V_s-
\min_{\bm\mu_A,\bm\mu_B}\frac{1}{2}\ln(\det\tilde{\bm\sigma}_{A,L})
\end{align}
for the first strategy, and
\begin{align}
I_J(\bm X_s,\bm Y_e)\simeq
\ln V_s
-\min_{\bm\mu_J}\frac{1}{2}\ln(a_{J1} a_{J2})
=\ln V_s-\min_{\bm\mu_J}\frac{1}{2}\ln(\det\tilde{\bm\sigma}_{A,J})
\end{align}
for the second strategy.  In the limit $V_s\rightarrow\infty$, the difference between these two mutual informations is equal to the OGD of $\rho_{AB}$,
\begin{equation}
I_J(\bm X_s,\bm Y_e)-I_L(\bm X_s,\bm Y_e)=D_{\text{OGD}}(B\rightarrow A).
\end{equation}
This relation provides the operational significance for our measure.

For some Gaussian states the local and joint Gaussian measurements used to minimize the conditional entropies for the OGD of ${\rho}_{AB}^{\prime}$ are independent of the value $V_s$, as shown for the states considered in Sec.~\ref{Asy-Sta}, whose conditional probability distributions are symmetric Gaussian functions.  In this case, one can easily see, for example, by considering Eqs.~(\ref{Asy-Loc-CM}) and~(\ref{Asy-Joi-CM}), that
\begin{align}
I_J(\bm X_s,\bm Y_e)-I_L(\bm X_s,\bm Y_e)
=D_{\text{OGD}}(B\rightarrow A)-D_{\text{OGD}}^{\prime}(B\rightarrow A),
\end{align}
where $D_{\text{OGD}}^{\prime}(B\rightarrow A)$, the OGD for the state ${\rho}_{AB}^{\prime}$ after encoding, is zero for maximal encoding.  According to this relation, the difference between mutual informations obtained by the above strategies is equal to the amount of nonclassical correlation in terms of OGD consumed by encoding the signal. This implies that, for any value of $V_s$, there is no difference between the two strategies for the CC state, $I_L(\bm X_s,\bm Y_e)=I_J(\bm X_s,\bm Y_e)$; however, for the CA state the joint Gaussian measurements strategy is always advantageous with respect to the local Gaussian measurements strategy, $I_L(\bm X_s,\bm Y_e)<I_J(\bm X_s,\bm Y_e)$.

\section{Conclusion}
\label{sec:discussion}

We have shown that operational Gaussian discord (OGD) is a new discord-type measure of nonclassical correlations for Gaussian states, which can be experimentally measured by using local and joint Gaussian measurements. We have demonstrated an operational significance for this measure in terms of a Gaussian quantum protocol for extracting information about a classical signal encoded on one subsystem; for maximal encoding, OGD is the additional accessible information that comes available when an  experimentalist throws off the shackles of local Gaussian measurements and starts using joint Gaussian measurements.  This measure might also be useful for quantifying nonclassical correlations in resources of other Gaussian protocols that involve Gaussian states, Gaussian operations, and Gaussian measurements.

An interesting open question is how to define a similar measure for discrete-variable systems.  This measure can be defined as the difference between two conditional entropies of one subsystem minimized by local and joint measurements. Such a measure might have an operational significance in terms of the discrete-variable version of the protocol we considered in this paper and other quantum protocols.


\ack
This work was supported in part by the Australian Research Council Centre of Excellence for Quantum Computation and Communication Technology (Project No.~CE110001027) and by U.S.\ National Science Foundation Grant Nos.~PHY-1314763 and PHY-1212445.
\appendix
\section*{Appendix. Gaussian measurements and entropy of Gaussian probability distributions}
\label{app}

In quantum mechanics, the uncertainty principle only allows noisy simultaneous measurements of the phase-space quadratures~\cite{Ulf}. In the general formalism for phase-space measurements, the POVM elements associated with the outcomes $(x,p)$ of a single-mode measurement are given by
\begin{equation}
\Pi(x,p)=\frac{1}{4\pi} D(x,p)\Pi_{0}D^{\dagger}(x,p),
\end{equation}
where $D(x,p)$ is the displacement operator, $\int d x\, d p\, \Pi(x,p)=\mathbb{1}$, and the quantum state $\Pi_{0}$, which can be assumed to have zero first-order moments, is a characteristic of the measurement device, sometimes called a quantum filter or ruler~\cite{Wod,Buzek}.  The phase-space measurements for which $\Pi_{0}$ is a Gaussian state are called Gaussian measurements~\cite{Weedbrook}; we show in the following that the outcome probability distributions are Gaussian.  It was shown, in general, that Gaussian operations can be implemented using linear-optical elements and homodyne measurements~\cite{Cirac02,Eisert03}.  Hence, Gaussian measurements are equivalently defined as a set of measurements that can be implemented using Gaussian ancilla states, Gaussian unitary operations, and homodyne measurements~\cite{FM,TS}.  As discussed in Sec.~\ref{sec:GQD}, we are interested in rank-one Gaussian measurements, i.e., $\Pi_{0}$ is a squeezed-vacuum state, as these states satisfy the minimum uncertainty relation, and the measurement is as accurate as possible.

Note that Gaussian measurements can be put in a slightly more general framework that uses  measurements with noncovariant POVMs $\Pi(x,p,y)=p(y)D(x,p)\Pi_{0}(y)D^{\dagger}(x,p)/{4\pi}$, where the Gaussian states $\Pi_{0}(y)$ are labeled by an additional parameter $y$, and $p(y)$ is the probability of outcome $y$, since $\int d x\, d p\,\Pi(x,p,y)=p(y)\mathbb{1}$~\cite{Paris10}. These measurements can be thought of as performing one of a set of random Gaussian measurements by flipping a coin, governed by the probability $p(y)$, to choose which Gaussian measurement. Since we optimize over all Gaussian measurements, a random choice of Gaussian measurements would not be optimal.  Formally, including the parameter $y$ in the minimization of the entropy would pick out the value of $y$ for the optimal measurement and lead to the same minimum entropy.

As discussed in the main text, a two-mode Gaussian measurement can be implemented using linear-optical elements and single-mode Gaussian measurements; the POVM elements are
\begin{equation}
\label{2POVM}
\Pi(\bm X)=\frac{1}{16\pi^2}D(\bm\xi)\Pi_{0}D^{\dagger}(\bm\xi).
\end{equation}
Here $\Pi_0$ is a Gaussian state with zero mean quadratures and the covariance matrix $\bm\mu_J$ of Eq.~(\ref{Joi-Cov}); $\bm X=\langle\hat{\bm X}\rangle=(x_A,p_A,x_B,p_B)$ is the vector of mean quadratures of the state $\Pi(\bm X)$ and represents the outcomes of the measurement; $\bm\xi=(\xi_1,\xi_2,\xi_3,\xi_4)\in \mathbb{R}^4$ and $\bm\xi =-\bm X\bm J/2$ with
\begin{equation}
\bm J=
\begin{pmatrix}
\bm J_A&0\\
0&\bm J_B
\end{pmatrix}
=
\begin{pmatrix}
0&1&0&0\\
-1&0&0&0\\
0&0&0&1\\
0&0&-1&0
\end{pmatrix}
\end{equation}
being the fundamental symplectic matrix; and
\begin{equation}
D(\bm\xi)=e^{i\bm\xi\hat{\bm X}^T}=e^{-i\bm X\bm J\hat{\bm X}^T/2}
=e^{i(p_A\hat{x}_A-x_A\hat{p}_A+p_B\hat{x}_B-x_B\hat{p}_B)/2}
=D(\bm X)
\end{equation}
is the two-mode displacement operator.

By using POVM elements (\ref{2POVM}) The probability distribution of the outcomes of the Gaussian measurement performed on a bipartite quantum system in state $\rho_{AB}$ is given by
\begin{equation}
P(\bm X)=\text{Tr}[\rho_{AB}\Pi(\bm X)].
\end{equation}
By using the characteristic function of the state, $\Phi_{AB}(\bm\xi)=\text{Tr}[\rho_{AB}D(\bm\xi)]$, and the characteristic function of the POVM-element states, $\Phi_{\Pi(\bm X)}(\bm\xi)=\exp(-\frac12\bm\xi\bm\mu_J\bm\xi^{T}+i\bm\xi\bm{X}^T)/(16\pi^2)$, this distribution can be written as
\begin{equation}
P(\bm X)=\int\frac{\bdif\xi}{\pi^2}\,\Phi_{AB}(\bm\xi) \Phi_{\Pi(\bm X)}(-\bm\xi).
\end{equation}
For a zero-mean Gaussian state $\rho_{AB}$ with covariance matrix $\bm\sigma_{AB}$, the characteristic function becomes $\Phi_{AB}(\bm\xi)=\exp(-\frac12\bm\xi\bm\sigma_{AB}\bm\xi^{T})$, so we have
\begin{equation}
\label{pro-dis}
P(\bm X)
=\frac{1}{16\pi^4}\int\bdif\xi\,e^{-\bm\xi(\sigma_{AB}+\bm\mu_J)\bm\xi^T/2} e^{-i\bm\xi\bm{X}^T}
=\frac{e^{-\bm{X}\tilde{\sigma}_{AB}^{-1}\bm{X}^T/2}}{4\pi^2\sqrt{\det\tilde{\bm\sigma}_{AB}}}.
\end{equation}
Thus the probability distribution is a Gaussian function with covariance matrix $\tilde{\bm\sigma}_{AB,J}=\bm\sigma_{AB}+\bm\mu_J$, as in Eq.~(\ref{CM-JM}).

Using the continuous Shannon (differential) entropy, the entropy of the joint probability distribution can be found as
\begin{align}\label{HAB}
H(\tilde A, \tilde B)=-\int\bdif X P(\bm X) \ln(P(\bm X))
=\frac{1}{2}\ln(\det \tilde{\bm\sigma}_{AB,J})+2\ln(2\pi e).
\end{align}
The constant $2\ln(2\pi e)$ does not have an absolute significance; the continuous entropy is only defined up to an additive constant.  The difference between two such entropies does, however, have an absolute significance.

When the covariance matrix $\tilde{\bm\sigma}_{AB,J}$ is written in terms of the block matrices $\tilde{\mathbf{A}}$, $\tilde{\mathbf{B}}$, and $\tilde{\mathbf{C}}$ of Eq.~(\ref{CM-JM}), the inverse is given by~\cite{MatMat}
\begin{equation}
\tilde{\bm\sigma}_{AB,J}^{-1}=
\begin{pmatrix}
\tilde{\bm\sigma}_{A,J}^{-1} & -\tilde{\bm\sigma}_{A,J}^{-1}\tilde{\mathbf{C}}\tilde{\mathbf{B}}^{-1} \\
-\tilde{\bm\sigma}_{B,J}^{-1}\tilde{\mathbf{C}}^T\tilde{\mathbf{A}}^{-1}&
\tilde{\bm\sigma}_{B,J}^{-1}
\end{pmatrix}
=
\begin{pmatrix}
\tilde{\bm\sigma}_{A,J}^{-1} & -\tilde{\mathbf{A}}^{-1}\tilde{\mathbf{C}}\tilde{\bm\sigma}_{B,J}^{-1} \\
-\tilde{\mathbf{B}}^{-1}\tilde{\mathbf{C}}^T \tilde{\bm\sigma}_{A,J}^{-1}&
\tilde{\bm\sigma}_{B,J}^{-1}
\end{pmatrix},
\end{equation}
where
\begin{equation}
\tilde{\bm\sigma}_{A,J}=\tilde{\mathbf{A}}-\tilde{\mathbf{C}}\tilde{\mathbf{B}}^{-1}\tilde{\mathbf{C}}^{T}, \ \ \ \ \ \ \ \ \ \ \
\tilde{\bm\sigma}_{B,J}=\tilde{\mathbf{B}}-\tilde{\mathbf{C}}^T\tilde{\mathbf{A}}^{-1}\tilde{\mathbf{C}}.
\end{equation}
By using this expression and the probability distribution~(\ref{pro-dis}), one can easily find the conditional probability distribution for $A$, given measurement results on~$B$, as
\begin{equation}\label{PAgivenB}
P(x_A,p_A|x_B,p_B)=P(\bm X_A|\bm X_B)
=\frac{e^{-\frac{1}{2}\bm{R}\tilde{\bm\sigma}_{A,J}^{-1}\bm{R}^T}}{2\pi\sqrt{\det\tilde{\bm\sigma}_{A,J}}},
\end{equation}
where $\bm R=\bm X_A-\bm X_B\tilde{\bm B}^{-1}\tilde{\bm C}^T$.  The covariance matrix of this distribution, $\tilde{\bm\sigma}_{A,J}$, is independent of the outcomes $x_B$ and $p_B$. Hence, the continuous Shannon entropy of the conditional probability distribution is given by
\begin{equation}\label{HAgivenB}
H(\tilde A|\tilde B)=\frac{1}{2}\ln(\det\tilde{\bm\sigma}_{A,J})+\ln(2\pi e).
\end{equation}

The conditional entropy~(\ref{HAgivenB}) can also be written as
\begin{equation}
H(\tilde A|\tilde B)=\frac{1}{2}\ln\left(\frac{\det\tilde{\bm\sigma}_{AB,J}}
{\det \tilde{\mathbf{B}}}\right)+\ln(2\pi e),
\end{equation}
since for joint classical probability distributions we have $H(\tilde A|\tilde B)=H(\tilde A,\tilde B)-H(\tilde B)$.  This relation can be regarded as a consequence of the identities
\begin{equation}
\frac{\det\tilde{\bm\sigma}_{AB,J}}
{\det\tilde{\mathbf{B}}}=
\det\tilde{\mathbf{A}}\left(1
-\text{Tr}[\tilde{\mathbf{C}}\tilde{\mathbf{B}}^{-1}\tilde{\mathbf{C}}^T\tilde{\mathbf{A}}^{-1}]
+\frac{(\det\tilde{\mathbf{C}})^2}{\det\tilde{\mathbf{B}}}\right)
=\det\tilde{\bm\sigma}_{A,J}.
\end{equation}

We note that by setting $\bm\mu_J=0$, i.e., by removing the tildes on all quantities so that $\tilde{\bm\sigma}_{AB,J}=\bm\sigma_{AB}$, the probability distribution $P(\bm X)$ of Eq.~(\ref{pro-dis}) becomes the Wigner function of the joint state~$\rho_{AB}$, with Eq.~(\ref{HAB}) giving the continuous Shannon entropy of the Wigner function.  Moreover, we can model a local measurement on $B$ by removing the tildes from $\mathbf{A}$ and $\mathbf{C}$, i.e., by setting $\bm\mu_{A,J}=0=\bm\mu_{C,J}$ so that $\tilde{\bm\sigma}_{A,J}=\bm\sigma_{A}$ [see Eq.~(\ref{Con-CM})]; in this case the conditional distribution~(\ref{PAgivenB}) becomes the conditional Wigner function of $A$, with Eq.~(\ref{HAgivenB}) giving the corresponding continuous Shannon entropy.


\end{document}